\documentclass{article}

\usepackage[square,numbers]{natbib}
    \usepackage[final]{neurips_2023_ml4ps}

\usepackage{graphicx}
\usepackage[utf8]{inputenc} 
\usepackage[T1]{fontenc}    
\usepackage{hyperref}       
\usepackage{url}            
\usepackage{booktabs}       
\usepackage{amsfonts}       
\usepackage{nicefrac}       
\usepackage{microtype}      
\usepackage{xcolor}         

\title{Rare Galaxy Classes Identified In Foundation Model Representations}

\author{%
  Mike Walmsley\thanks{Second affiliation: Jodrell Bank Centre for Astrophysics, Department of Physics \& Astronomy, University of Manchester, Manchester M13 9PL, UK}\\
  Dunlap Institute for Astronomy and Astrophysics\\
  University of Toronto\\
  Toronto, ON M5S 3H4, Canada\\
  \texttt{m.walmsley@utoronto.ca} \\
  \And
  Anna M.M Scaife \\
  Jodrell Bank Centre for Astrophysics\\
  Department of Physics \& Astronomy\\
  University of Manchester\\
  Manchester M13 9PL, UK \\
  \texttt{anna.scaife@manchester.ac.uk} \\
}

\begin{document}

\maketitle

\begin{abstract}
We identify rare and visually distinctive galaxy populations by searching for structure within the learned representations of pretrained models. We show that these representations arrange galaxies by appearance in patterns beyond those needed to predict the pretraining labels. We design a clustering approach to isolate specific local patterns, revealing groups of galaxies with rare and scientifically-interesting morphologies.
\end{abstract}

\section{Introduction}
\label{introduction}

Astronomers often focus on specific classes of galaxy that have been particularly affected (or unaffected) by a process of interest \citep{Simmons2013, Grouchy2006, Fitts2018, Martin2019low}.
These classes may be extremely rare; `jellyfish' galaxies, for example, are only reliably identified in dense cluster environments \cite{McPartland2016, Bellhouse2022}.

It has recently become possible to automatically search for rare galaxy classes by manually identifying a small set of examples and then finetuning a pretrained model to search for more examples at scale. The most frequently-used galaxy model library is Zoobot \citep{Walmsley2023zoobot}, a set of models trained on 100M responses by Galaxy Zoo volunteers \citep{Masters2019a} answering a diverse set of classification tasks. 
By jointly predicting responses to all of these tasks, Zoobot's encoder learns to assign visually similar galaxies to similar points in representation space \citep{Walmsley2022practical}.
One can then straightforwardly (e.g. with a linear head) identify areas of representation space containing known examples of the target rare galaxy class.

But what if there are galaxy classes which are so rare we don't know to look for them?
Here, we reverse the finetuning workflow and instead identify rare classes of galaxies directly from Zoobot's representation, without any known examples.

\citet{Walmsley2022practical} suggested that the representation learned by Zoobot may arrange galaxies in patterns beyond those required to predict the supervised labels. 
Our approach here will both prove this thesis and demonstrate that, beyond being grouped by similarity, variations in the density of galaxies within that representation can be exploited to reveal new subclasses.

Our aim of finding rare galaxy classes is related to anomaly-finding.
Unlike standard anomaly-finding  \citep{Henrion2013, Giles2019, StoreyFisher2021}, we explicitly aim to identify anomalies with multiple examples; a one-off strange galaxy is a puzzle, but a set of them is a new population.

Our aim is also related to the topic of data-driven galaxy classification schema.
Astronomers often divide all galaxies into broad (i.e. not rare) classes and then compare their features. 
The most `meaningful' criteria with which to define these broad classes is a matter of decades-long debate \citep{Sandage1975}. Much work has been done to uncover alternative data-driven criteria both before \citep{Whitmore1984} and with \citep{Hocking2018,Cheng2021unsupervised} machine learning, but these alternative criteria are rarely used in practice. 
Our method is not intended to describe galaxies in general - less than 2\% of all galaxies studied here fall in one of our classes - but rather aims to isolate rare and visually distinctive classes within the bulk sample.

\section{Data}
\label{sec:pretrained_models}

We begin with the public pretrained models available from the Zoobot model library. We run all experiments with both the EfficientNetB0 and MaxViT Tiny  architectures (224px, three-channel) to test if adding multi-axis attention blocks alters the representation learned \citep{Tan2019a,Wightman2019,Tu2022}. These models have been referred to \citep{Walmsley2023zoobot} as galaxy `foundation' models, in that they are pretrained on a diverse task and extensive training data in order to learn an adaptable representation, but we highlight that have relatively few parameters; EfficientNetB0 and MaxViT Tiny are both low-parameter variants of their respective model families.
\label{sec:galaxy_images}

We construct our initial representations by applying the pretrained MaxViT Tiny and EfficientNetB0 Zoobot models to images from the DESI Legacy Surveys \citep{Dey2018}.
Our base catalog is the 8.7M galaxies of suitable brightness ($r < 19.0$) and angular extent (Petrosian radius $> 3$'') to have well-resolved features \cite{Walmsley2023desi}.

We use the predicted morphology labels (from the EfficientNetB0 model, for consistency) to automatically remove featureless (`Smooth' response probability $\geq 50$\%) and edge-on (`edge-on disk' $\geq 30$\%) galaxies before applying our search method. 
Naively applying our search method to this sample returns many clusters of artifacts (e.g. images contaminated by telescope issues).
We are primarily interested in rare galaxies, not rare image issues, so (outside of a final experiment, Sec. \ref{sec:results}) we remove these (`artifact $\geq$ 30\%) but note that flagging contaminated images may be a further useful application of our search method. Our final sample is 632k DESI Legacy Survey images of featured galaxies.

\section{Methods}

The representation dimensionality of the published MaxViT Tiny and EfficientNetB0 architectures is 1024 and 1280, respectively. Motivated by a PCA analysis of the representation variance in the original networks, showing that the representations are highly redundant, we find we can retrain these networks using a lower representation dimension of 128 without any measurable performance loss. We refer to our adjusted networks as `bottlenecked' networks.

We then apply dimensionality reduction to further compress our representation before searching for clusters in the compressed space. 

Clustering within the representation is non-trivial.
Supervised representations of galaxies have previously been noted as broadly smooth and naive clustering as unsuccessful \cite{Walmsley2022practical}.
We hypothesise that this global smoothness masks small-scale detailed structure and design our clustering approach to identify this structure.

We first reduce the bottlenecked representation dimensionality  with UMAP \cite{McInnes2018}. 
We then construct a tree of candidate clusters with HDBSCAN \citep{McInnes2017} and post-process that tree for our final clusters.

UMAP's \texttt{n\_neighbours} sets the local connectivity of the force-directed graph.
Each point (node) has edges to the closest \texttt{n\_neighbours}, and those edges apply forces weighted by distance. 
To find rare subclasses, \texttt{n\_neighbours} must be large enough that small clusters can form but small enough that those clusters are not `torn apart' by numerous non-cluster neighbours. We find \texttt{n\_neighbours}$=50$ provides a good balance for our goal of finding rare galaxy subtypes.

UMAP's \texttt{n\_components} sets the final dimensionality and hence the typical separation of galaxies. 
Reducing to $D = 2$ preserves only the most noteable clusters (though it is useful as a visualisation aid). Reducing to $D = 15$ leads to qualitatively less persuasive clusters, likely because of sparsity\footnote{A $15$D latin hypercube of 630k galaxies would have approx. 2.4 galaxies per dimension.}. We find $D=5$ provides a good balance for our dataset size.

We next apply HDBSCAN to identify clusters in the reduced representation.
HDBSCAN is typically used to identify well-separated clusters of significant size. 
To identify only confident clusters, we adjust the denoising distance transformation (`core distance') by setting \texttt{min\_samples}$=300$. We also set \texttt{min\_cluster\_size}$=10$ to allow for rare classes. To allow for variable size clusters, we define clusters as the leaf nodes in HDBSCAN's condensed tree rather than the default size-focused (`excess of mass`) node selection.

This configuration is able to identify small clusters within the smooth bulk sample, but at the cost of further splitting larger clusters. A hyperparameter search identifies no single $\lambda$ or $\epsilon$ threshold that selects small clusters without splitting larger clusters, and so instead we simply note that manually grouping visually similar clusters is trivial if required (all clusters presented here are shown as-found without manual grouping).

\section{Results}
\label{sec:results}

\begin{figure*}[hp]
    \centering
    \includegraphics[width=1.\textwidth]{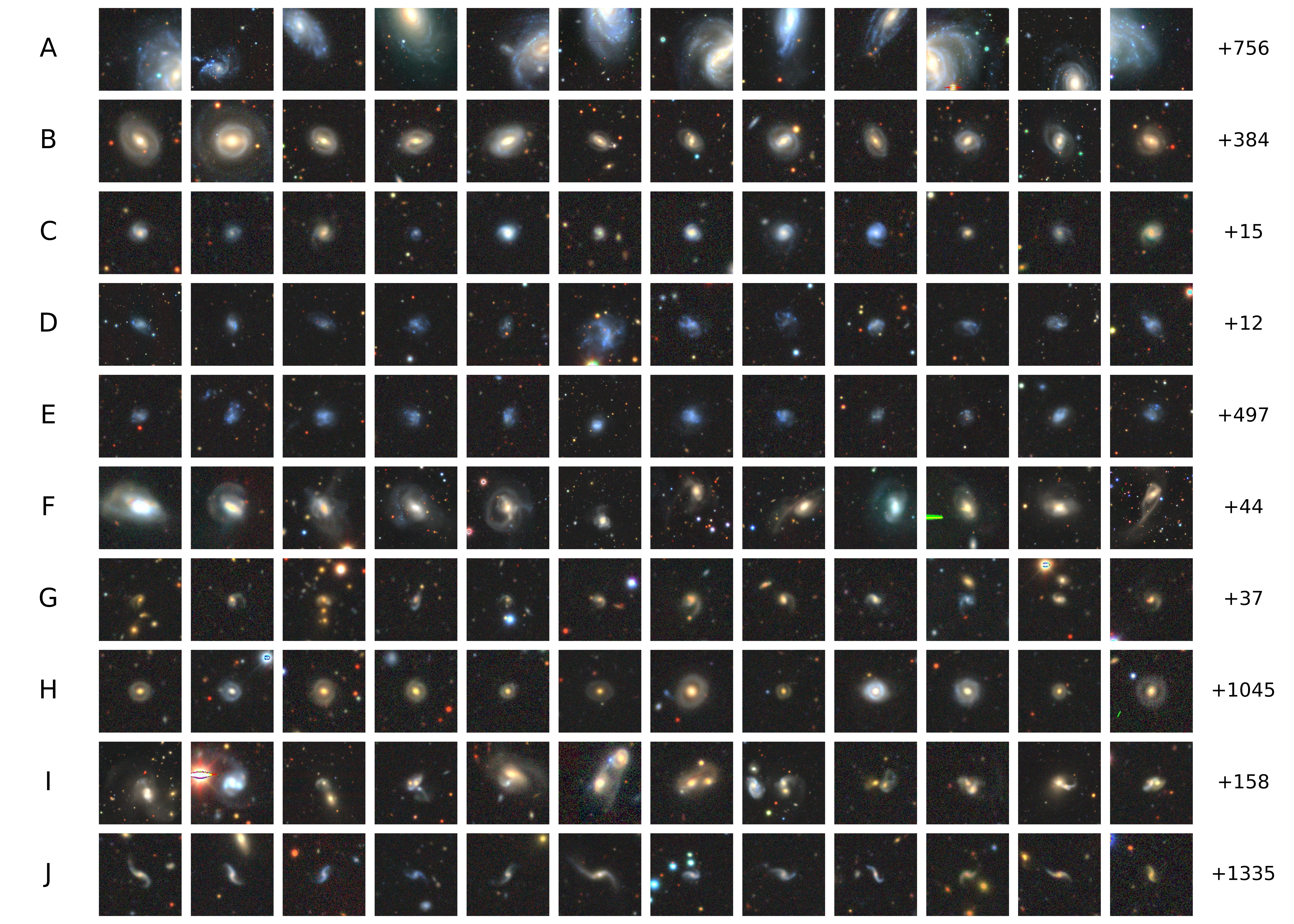}
    \caption{
        Selected galaxy classes identified within Zoobot's representation (EffNetB0 architecture). Notable examples include low-mass barred (D) and unbarred (E) galaxies, galaxies with clear low surface brightness structure (F), one-armed spirals (G) and merging pairs (I). Class A is reveals a DESI pipeline issue where extended clumpy spirals are `shredded' into many sources.
        }
        
    \label{fig:selected_classes}
    
\end{figure*}

Fig \ref{fig:selected_classes} shows classes identified directly from Zoobot's representation using our search method.
Our classes reveal visually cohesive groups of galaxies with rare morphologies. Our data-driven search identifies these without expert astronomer direction.

Several of our classes have not previously been automatically detected or studied at scale. Most notably, class D shows a population of galaxies with dominant bars and irregular clumpy spiral structure (likely low-mass and starforming). 

The classes are split in ways not obviously motivated by the supervised labels. For example, with class D (described above), the Galaxy Zoo labels on which Zoobot was trained include bar and spiral arm measures but do not specify if those spirals are irregular or clumpy. Similarly, class A shows `shredded' spiral galaxies which have been mistakenly recorded by the DESI pipeline as several sources - a distinction not present in the Galaxy Zoo labels. 

\begin{figure*}[hp]
    \centering
    \includegraphics[width=1.\textwidth]{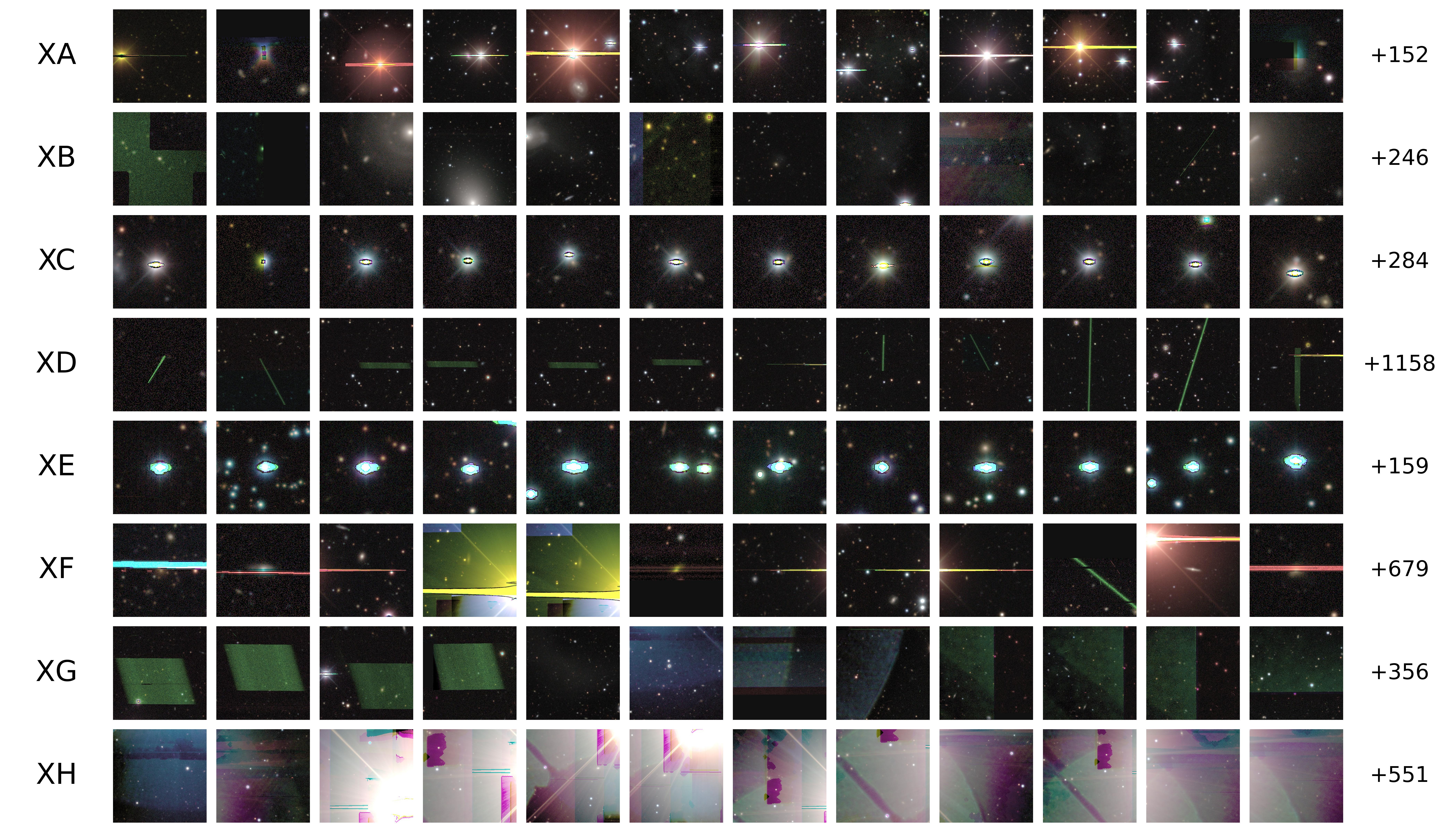}
    \caption{
        Selected artifact classes identified within Zoobot's representation (MaxViT architecture). Zoobot's representation places artifacts into visually distinct classes, even though Zoobot's pretraining labels do not distinguish between different types of artifact. }
    \label{fig:selected_artifacts}
\end{figure*}

To prove that the representation does indeed group galaxies independently of the training labels, we run our search method on our full sample prior to removing artifact images (Sec. \ref{sec:galaxy_images}). The response `Artifact' is a leaf in the Galaxy Zoo decision tree and so the labels include no further information on which type of artifact an image might be. Nonetheless, our search method reveals (Fig \ref{fig:selected_artifacts}) that Zoobot's representation is arranging artifacts by visual appearance. This directly demonstrates the internal emergence of classes beyond those needed for predicting training labels.

\section{Discussion}

Quantitative comparisons with existing methods are not possible because, to our knowledge, there are no previous works attempting to extract rare galaxy subclasses from a deep representations. Further, there is no commonly-accepted definition for what might constitute anomalies or anomaly subclasses. Standard benchmarks for this task (or for similarity metrics in general) would help advance the field.

Our representation is broadly resistant to clustering. Most galaxies (95\%+) cannot be confidently assigned to clusters, even with generous HDBSCAN mutual reachability distance settings. Adding a loss term to encourage preserving density \cite{Narayan2020.05.12.077776} further smooths the compressed representation. 
Whether this limited clustering is a consequence of the nature of supervised representations (and perhaps self-supervised representations e.g. \cite{Stein2022}) or induced by our compression process remains to be seen. Altering the training process (e.g. with a hybrid self/supervised approach, \cite{Walmsley2022Towards}) or the compression process might lead to more easily clustered representations. Alternatively, a continuous user-guided search algorithm may work better. 

\section{Conclusion }

We have proven that the representation learned by the Zoobot pretrained galaxy models groups visually similar galaxies together - even where those patterns are definitively independent of the pretraining labels. Further, we have shown that this grouping of similar galaxies leads to patterns (overdensities) which can be exploited to identify new subclasses. We applied UMAP and tuned HDBSCAN clustering to extract these patterns. Our clustering reveals groups of galaxies with rare and distinctive morphologies. Some of these groups have not previously been automatically identified or studied at scale; for example, low-mass barred galaxies with clumpy irregular spiral structure. We hope this simple search technique will help astronomers discover new unexpected populations of galaxies that shed light on the processes driving galaxy formation.

\begin{figure}[h!]
    \centering
    \includegraphics[width=0.5\textwidth]{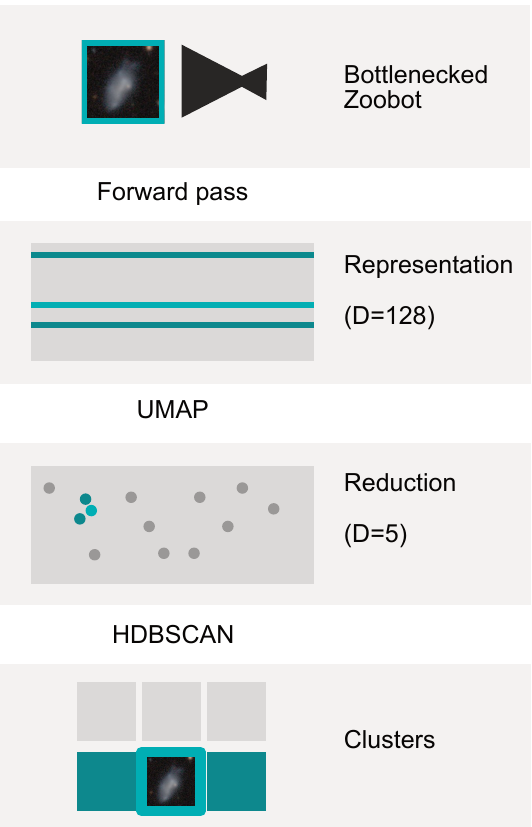}
    \caption{Schematic of search approach. Galaxy images are passed through pretrained Zoobot model, modified to use a D=128 feature bottleneck. Resulting representation is compressed via UMAP then clustered with HDBSCAN.}
    \label{fig:schematic}
\end{figure}

\bibliography{references}

\end{document}